\documentclass[12pt]{article}
\usepackage{putex}
\usepackage{graphicx}

\begin{document}

\preprint{PUPT-2394}

\institution{PU}{Joseph Henry Laboratories, Princeton University, Princeton, NJ 08544}

\title{Superluminal neutrinos and extra dimensions:\\ constraints from the null energy condition}

\authors{Steven S. Gubser}

\abstract{In light of the recent results from the OPERA collaboration, indicating that neutrinos can travel superluminally, I review a simple extra-dimensional strategy for accommodating such behavior; and I also explain why it is hard in this strategy to avoid violating the null energy condition somewhere in the extra dimensions.}

\date{September 2011}

\maketitle

The high statistical significance of the OPERA result \cite{OPERA:2011zb}, $(v_\nu - c) / c = (2.48 \pm 0.28 \, \hbox{(stat.)} \pm 0.30 \, \hbox{(sys.)}) \times 10^{-5}$, on the propagation of neutrinos with $\hbox{energy} \sim 17\,{\rm GeV}$, justifies some theoretical speculation, as well as demanding follow-up experimental work to confirm or refute one of the main consequences of special relativity.

\begin{figure}
  \centerline{\includegraphics[width=6.5in]{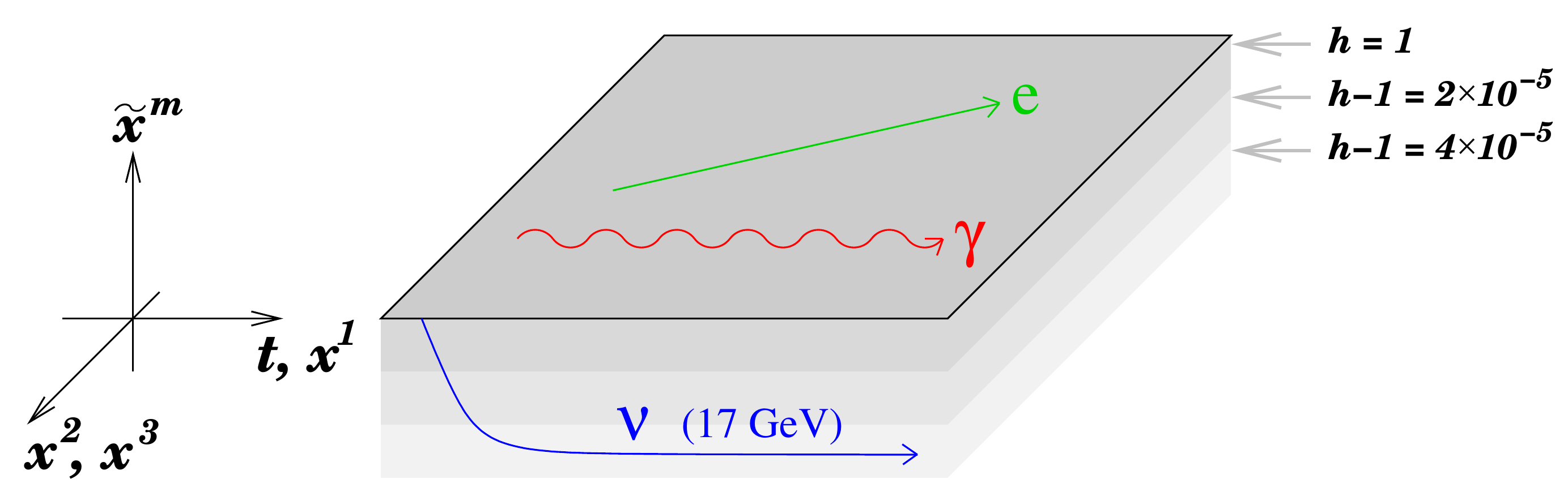}}
  \caption{A cartoon of a way in which extra dimensions can accommodate superluminal motion.  Photons and charged particles must be confined to a region where the blackening function $h$ is very close to $1$.  Superluminal particles (for example, $17\,{\rm GeV}$ neutrinos if the OPERA measurement is correct) explore regions of the extra dimensions where $h$ is larger than $1$.}\label{BLACKER}
 \end{figure}

There is already a considerable literature on how superluminal propagation could be accommodated in extra-dimensional scenarios, including the works \cite{Rubakov:1983bb,Kaelbermann:1998hu,Visser:1985qm,Chung:1999xg,Kraus:1999it,Youm:2001sw,Kiritsis:1999tx,Alexander:1999cb,Bowcock:2000cq,Csaki:2000dm,Cline:2001yt,Dubovsky:2001fj,Deffayet:2001aw,Frey:2003jq,Ganor:2006ub,Koroteev:2007yp,Gubser:2008gr}.\footnote{I do not attempt here to summarize other aspects of the large and diverse literature on superluminal motion.}  The approaches of these papers, adapted to the current circumstance, are typified by the following geometrical configuration.  Consider a $D$-dimensional line element
 \eqn{BulkGeometry}{
  ds_D^2 = e^{2A} (-h c^2 dt^2 + d\vec{x}^2) + e^{2B} d\tilde{s}_{D-4}^2
    \,,
 }
where $c$ is a constant, and $A$, $B$, and $h$ are dimensionless functions depending only on the $D-4$ coordinates of $d\tilde{s}^2_{D-4}$, which I will denote as $\tilde{x}^m$.  The coordinates $t$ and $\vec{x} = (x^1,x^2,x^3)$ parametrize non-compact four-dimensional spacetime.  Matter can be confined to branes which are at least partially localized within the extra dimensions, but which extend over the $t$ and $\vec{x}$ directions.  Assuming that causal propagation is limited to timelike and null trajectories with respect to the bulk geometry \eno{BulkGeometry}, the maximum speed of matter trapped on a brane at a specific pointlike location $\tilde{x}_*$ in the extra dimensions is $v_* = \sqrt{h(\tilde{x}_*)} c$.  We can then simply assume that photons propagate on a brane where $h=1$, so that $c$ is the speed of light, while neutrinos propagate in such a way as to explore regions of spacetime where $h$ is slightly larger than $1$.  See figure~\ref{BLACKER}.  This picture is already enough---in outline---to accommodate the OPERA results.
The function $h$ is sometimes called the blackening function, because in geometries with black brane horizons parallel to the $\vec{x}$ directions, $h$ vanishes at the event horizon.  Some of the works cited above employ black brane geometries, but this is not necessary; all that is needed is some slight variation of $h$ over the extra dimensions, together with a mechanism that allows neutrinos of the type measured in the OPERA experiment to explore ``less black'' regions (i.e.~regions with larger $h$) than those experienced by the photon.
 
One must keep in mind that there are stringent bounds on the propagation speed of lower energy neutrinos \cite{Hirata:1987hu,Bionta:1987qt,Longo:1987ub}.  Thus, one would have to imagine a situation where the wave-function for low-energy neutrinos (say $10\,{\rm MeV}$ if one is considering the bounds just cited) is mostly confined near the photons' brane (or at another location where $h$ is very nearly equal to unity), but when the neutrino energy is at the $17\,{\rm GeV}$ scale the wave-function has significant support at locations where $h$ is larger than $1$.  Meanwhile, the speed limit on electrons (even very energetic ones, at the scale of $50 - 100\,{\rm TeV}$) is much more stringent than for neutrinos \cite{Jacobson:2002hd}.  So electrons must presumably stay much closer to the photons' brane than neutrinos do.  This seems odd given that neutrinos and leptons form electroweak doublets.  However, an optimistic view is that it indicates that the extra-dimensional dynamics relevant to superluminal neutrino propagation is somehow tied to electroweak symmetry breaking.  If so, it is perhaps a new reason to think that extra dimensional or stringy physics may be accessible not too far above the electroweak scale---as suggested on other grounds in \cite{ArkaniHamed:1998rs,Antoniadis:1998ig}.

There is a serious obstacle to realizing geometries of the form \eno{BulkGeometry}: They typically violate the null energy condition at some point in the extra dimensions \cite{Cline:2001yt,Gubser:2008gr}.  Assessing the seriousness of this obstacle will be the focus of the rest of the current work.

The null energy condition is the requirement that $T_{MN} \xi^M \xi^N \geq 0$ for any null vector $\xi^M$, where $T_{MN}$ is the stress-energy tensor.\footnote{For a diagonal, isotropic stress-energy tensor, the null energy condition states that $\rho+p \geq 0$, where $\rho$ is the energy density and $p$ is the pressure.  It is the weakest of the commonly used positive energy conditions, which are needed in order to prove the positive mass and area increase theorems of general relativity \cite{Schoen:1979zz,Witten:1981mf,Hawking:1971tu}.}  Throughout, indices $M$ and $N$ indicate $D$-dimensional tensors.  Assuming Einstein's equations in $D$ dimensions,
 \eqn{EinsteinEqs}{
  R_{MN} - {1 \over 2} R g_{MN} = T_{MN} \,,
 }
an equivalent phrasing of the null energy condition is $R_{MN} \xi^M \xi^N \geq 0$.  Consider the choice $\xi^M = (e^{-A}/\sqrt{h},e^{-A},0,0,0^{(D-4)})$, where I have set $c=1$ for notational convenience and used $0^{(D-4)}$ to indicate $D-4$ copies of $0$.  Then $R_{MN} \xi^M \xi^N = -R^0_0 + R^1_1$.  By direct computation,
 \eqn{NullGeneral}{
  4h^2 e^{2B} (-R^0_0 + R^1_1) = \tilde\square \left( h^2 \right) - 
    3 \tilde{g}^{mn} \partial_m h \partial_n h + 
    8 \tilde{g}^{mn} \partial_m h \partial_n A + 
    2 (D-6) \tilde{g}^{mn} \partial_m h \partial_n B
    \,.
 }
The easiest way to see that \eno{NullGeneral} leads to problems is to consider what happens near a generic maximum of $h$---meaning a maximum about which the corrections to constant behavior start at quadratic order.  Also assume that $A$ and $B$ are non-singular at this maximum.  Then all terms on the right hand side of \eno{NullGeneral} vanish except the first, and this first term is negative.  In other words, the null energy condition gets violated at the location in the extra dimensions where the speed of light is maximized.

This argument seems to offer a few potential loopholes: For example, one can consider a non-generic maximum of $h$.  A complementary strategy, for $D \neq 6$, is based on the following integral inequality.  Set $B = {4 \over 6-D} A$: this can be achieved simply by redefining $d\tilde{s}_{D-4}^2$, and it causes the last two terms on the right hand side of \eno{NullGeneral} to cancel.  Now integrate \eno{NullGeneral} over the extra dimensions, and recall that the left hand side is non-negative.  The result of this integration is the inequality
 \eqn{hInequality}{
  \int d^{D-4} \tilde{x} \, \sqrt{\tilde{g}} \, \left[ 
    \tilde\square \left( h^2 \right) - 
      3 \tilde{g}^{mn} \partial_m h \partial_n h \right] \geq 0 \,,
 }
where $\tilde{g} = \det \tilde{g}_{mn}$.  The first term inside square brackets in \eno{hInequality} is a total derivative, so it vanishes when integrated over a compact space, assuming that the space has no boundaries.  The second term is non-positive, and its integral vanishes only if $h$ is a constant.  So the only way to satisfy the inequality in \eno{hInequality} is to have $h$ everywhere constant: in short, no variability of the speed of light over the extra dimensions is allowed.

There are still some potential loopholes.  The extra dimensional manifold may be non-compact; or it might have boundaries; or there may be some sort of localized singularities on it which invalidate the use of \eno{hInequality}.  I will now consider each of these possible loopholes in turn.

Non-compactness is limited by the need to have a normalizable graviton.  Consider a perturbed geometry
 \eqn{Graviton}{
  ds_D^2 = e^{2A} \left( -h dt^2 + d\vec{x}^2 + 2 \lambda e_{23} dx^2 dx^3 + 
   {\lambda^2 \over 2} e_{23}^2 \left[ (dx^2)^2 + (dx^3)^2 \right] \right) + 
   e^{2B} d\tilde{s}_{D-4}^2 \,,
 }
where $e_{23}$ depends on $t$, $x^1$, and $\tilde{x}^m$, $\lambda$ is a small parameter, and I have again set $c=1$ for convenience.  The Einstein-Hilbert lagrangian is
 \eqn{EHLag}{
  \sqrt{g} R &= {\cal O}(\lambda^0) + 
   \lambda^2 e^{4A + (D-4)B} \sqrt{h \tilde{g}} \left[ 
    {1 \over e^{2A} h} (\partial_t e_{23})^2 - 
     e^{-2A} (\partial_{x^1} e_{23})^2 - 
     e^{-2B} (\partial_m e_{23})^2
   \right]  \cr
    &\qquad\qquad{} + {\cal O}(\lambda^3) + \hbox{(total derivatives)} \,.
 }
The ${\cal O}(\lambda^0)$ term has to do only with satisfying the unperturbed Einstein equations and does not depend on $e_{23}$.  The total derivative terms integrate to extrinsic curvature terms on any boundaries and are also unimportant for present purposes.  In order for four-dimensional gravitons to have a finite norm, the term shown explicitly in \eno{EHLag} should be integrable across the extra dimensions.  I will not enter into a detailed analysis of the linearized equation of motion for $e_{23}$; however, because constant $e_{23}$ is a solution, a reasonable expectation is that the variation of $e_{23}$ across the extra dimensions is not dramatic when the wavelength and frequency in four dimensions are long compared to the size of the extra dimensions.  Thus a good guide to integrability is the requirement that $e^{2A+(D-4)B} \sqrt{h\tilde{g}}$ and $e^{2A+(D-4)B} \sqrt{\tilde{g}/h}$ are integrable with respect to the coordinate measure $d^{D-4} \tilde{x}$, where as usual $\tilde{g} = \det \tilde{g}_{mn}$.  This requirement forces us to discard many non-compact geometries, like the $AdS_5$-Schwarzschild metric, in which $e^{2A}$ becomes large as $h$ approaches a maximum.

The trouble with boundaries is that one or more of them (namely, the one(s) where $h$ reaches a maximum) will need to have a source of stress-energy localized on it which is itself in violation of the null energy condition if \eno{hInequality} is satisfied in the bulk.  A closely related line of argument is that $-T^0_0 + T^1_1$ must have contributions localized at the boundaries in order to satisfy appropriate boundary conditions on Einstein's equations; then the sum of such contributions should be added to the bulk integral in \eno{hInequality} in such a way that the total is the integral of $4h^2 e^{2B} (-T^0_0 + T^1_1)$, including distributional terms at the boundaries.  The non-negativity of this improved integral, together with the requirement that both the continuous and distributional contributions to the stress tensor satisfy the null energy condition, leads quickly to the conclusion that $h$ must be constant.

It is hard to give a uniform treatment of singularities in the extra dimensions, because when curvatures diverge the Einstein equations are probably not a reliable guide to the physics.  A criterion I have espoused in the past \cite{Gubser:2000nd} is that naked curvature singularities should be allowed in extra-dimensional constructions when they can be realized as limits of geometries which have an event horizon and are smooth outside it.  Singularities where $h$ becomes large, or where it reaches a maximum, obviously cannot be so realized without violations of the null energy condition outside the horizon.  The reason is simply that $h \to 0$ at a horizon, so a black brane which is very close to the desired geometry would have $h$ first rising to a maximum, then falling to $0$ as it approaches the horizon.  The maximum would be a generic one (at least for most temperatures), so we're back to an obvious violation of null energy.  The criterion of \cite{Gubser:2000nd} is not sharp: that is, physically allowed singularities, like D8-branes, exist which violate it.  A closely related criterion was suggested in \cite{Maldacena:2000mw}: the magnitude of the metric component $g_{00}$ must remain bounded above as one approaches the singularity (weak form), or it must not increase as one approaches the singularity (strong form).  The strong form again rules out D8-branes.  Insofar as we are opposed to permitting extra-dimensional geometries that allow superluminal propagation, it seems sensible to suggest a new variant of this criterion, for metrics of the form \eno{BulkGeometry}: $h$ must not increase as one approaches a singularity.

It is intriguing that the argument based on the integral inequality \eno{hInequality} is unavailable when $D=6$.  It is therefore natural to consider more closely in $D=6$ the possibility of non-generic maxima of $h$.

To summarize: While it is easy to construct local models where extra-dimensional metrics of the form \eno{BulkGeometry} allow superluminal propagation, the null energy condition makes it hard to embed these local models into a compactification with reasonable properties, for example the existence of four-dimensional gravity.  The difficulties tend to arise especially at the location in the extra dimension where the propagation speed is the fastest.  Efforts to escape these difficulties, for example by supposing that the propagation speed is unbounded above, or that it is bounded but the maximum is not attained, have not led me so far to viable constructions which avoid explicit violations of the null energy condition.

\section*{Acknowledgments}

This work was supported in part by the Department of Energy under Grant No.~DE-FG02-91ER40671.

\bibliographystyle{ssg}
\bibliography{time}
\end{document}